# Does a particle swept by a turbulent liquid diffuse?

Moshe Schwartz

School of Physics and Astronomy,
Raymond and Beverly Faculty of Exact Sciences,
Tel Aviv Univrersity, 69978 Tel Aviv, Israel

Gad Frenkel
Earth Science and Engineering,
Imperial College, London SW7 2AZ, UK

S. F. Edwards

Cavendish Laboratory,
Madingley Rd. CB3OHE Cambridge, UK

Abstract

Since the famous 1926 paper by Richardson the relative diffusion of two particles in a turbulent liquid has attracted a lot of interest. The motion of a single particle on the other hand is usually considered not to be especially interesting. The widely accepted picture is that the velocity of the particle has short-range correlations in time, resulting in motion that is diffusive on time scales large compared to the correlation time. We find, however, that the correlation time is infinite and that the mean square displacement (MSD), $\Delta^2$, is not linear in the traversed time, $T$, which would correspond to diffusion, but rather $\Delta^2 \propto T^{6/5}$ for large times. Namely, the motion is slightly super diffusive. The conditions, which will allow observing this behavior, are discussed.

The Richardson 1926 paper [1] on the relative diffusion of two particles in the atmosphere had been the precursor and trigger of the enormous work done on turbulence, starting with the work of Kolmogorov [2,3] and Obukhov [4] in the early forties of the 20[th] century and continuing ever since. The same experiment is also the corner stone of the more modern field of passive scalar turbulence [5,6], which developed within the general field of turbulence, with the relative diffusion of two particles as one of its focal points. Relative diffusion characterizes the spread of a smoke plume carried by turbulent flow. The diffusion of a single particle on the other hand defines some kind of a boundary within which the plum is effectively confined. This is clearly not less important than the spread for assessing the risks of pollutants in the atmosphere or in the ocean. The diffusion of a single particle has not attracted much interest, however, because its statistics seems to have been adequately described by an old argument of G. I. Taylor [7], who concluded that a single particle diffuses on large time scales, while at shorter times it might be ballistic. The Taylor argument is based on the assumption that the velocities of a particle swept by a liquid have a finite correlation time, $\tau_C$. Consequently, a simple order of magnitude calculation gives the effective diffusion constant of the particle as

$$D = \langle \mathbf{v}^2 \rangle \tau_C, \tag{1}$$

where $\mathbf{v}$ is the velocity of the particle, $\langle \cdots \rangle$ denotes an average over all possible trajectories, and the frame of reference is chosen to render the average velocity zero. The only possible weakness in the above is the assumption of a finite correlation time, $\tau_C$. A possible origin for an infinite correlation time of the Lagrangian velocity of the particle is the behavior of the long range disturbances of the velocity field, driving the particle, that die out on time scales that diverge with the typical size of the disturbance. An infinite correlation time leads to super-diffusion, which is characterized by the fact that the MSD is a power law in the elapsed time with a power which is larger than one [8-11]. We check, in the following whether super-diffusion can happen indeed with ordinary turbulence.

An ordinary turbulent incompressible liquid can be described by the noise driven Navier Stokes equation [12],

$$\partial_t \mathbf{V} + \mathbf{V} \cdot \nabla \mathbf{V} = -\nabla P + \nu \nabla^2 \mathbf{V} + \eta \tag{2}$$

and the incompressibility condition

$$\nabla \cdot \mathbf{V} = 0. \tag{3}$$

The spatial Fourier transform of the noise, $\eta(\mathbf{q},t)$, is taken to have zero mean and to have the following correlations

$$\langle \eta_i(\mathbf{q},t) \eta_j(\mathbf{p},t') \rangle = \delta_{ij} D(q) \delta(\mathbf{q}+\mathbf{p}) \delta(\mathbf{t}-\mathbf{t}'). \tag{4}$$

One of the objects of the theory of turbulence is to obtain the steady state time dependent velocity correlations,

$$\langle \mathbf{V}_i(\mathbf{q},t'+t) \mathbf{V}_j(\mathbf{p},t') \rangle = [\delta_{ij} - \frac{q_i q_j}{q^2}] \delta(\mathbf{q}+\mathbf{p}) \Phi(q,t), \tag{5}$$

which depend on the noise characteristics.

Ordinary turbulence is characterized by three well separated "momentum" ranges. The first is the range of wave vectors in which the noise pumps energy into the system. The upper limit of that range is $q_0$ above which the noise correlation, $D(q)$, vanishes effectively. The inertial range is bound from below by $q_0$ and from above by $q_D = \frac{\langle \mathbf{V}^2 \rangle^{1/2}}{\nu}$, which indicates the "momentum" above which dissipation of mechanical energy into heat due to viscosity becomes effective. The usual situation is that $\frac{q_D}{q_0} \gg 1$ (figure 1). The last relation is the one that allows observation of the characteristic exponents describing steady state turbulence. Since that was always the region of interest in the study of turbulence, it was only natural that the study of passive scalars will concentrate on that region. Actually, some work went beyond that and included also the dissipative region [13] but consideration of velocity correlations within the dissipative and inertial ranges for the motion of a particle swept by a turbulent liquid are only useful for times, which are not too long. The long time behavior of the trajectory of such a particle involves also very large spatial scales and is consequently mostly affected by very large scale disturbances of the velocity field. Namely, the "momentum" range relevant to the asymptotic behavior of the MSD, $\Delta^2$, is not the inertial range but rather the energy pumping range. To get an idea of the relevant velocity correlations in that range, consider first a different kind of turbulence in which the liquid is driven by spatial white noise described by taking $D(q)$ in equation (4) to be a constant in $q$. Some years ago [14] two of us (M.S and S.F.E) obtained the time dependent correlations of the Fourier components of the velocity field, $V_i(\mathbf{q},t)$, for an incompressible liquid driven by spatial white noise (SWN) for small $q$ and large times,

$$\Phi(q,t) = Aq^{-5/3}\exp[-Bqt^{3/5}]. \qquad (6)$$

(Note that the factor $q^{-5/3}$ may cause some confusion, because it is exactly the exponent appearing in the power spectrum of Kolmogorov turbulence and hence equation (6) may be mistaken to be the Kolmogorov power spectrum. The equal time correlations behave indeed as $q^{-5/3}$ in the case of SWN, but these are correlations of the Fourier components of the velocity field and not the power spectrum.)

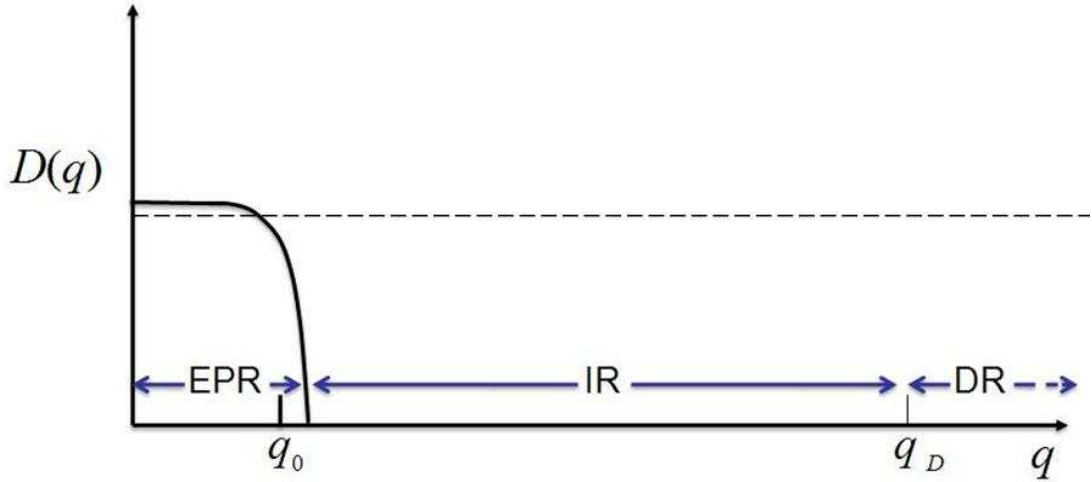

Figure 1: Driving noise correlations for ordinary turbulence (solid line) and SWN turbulence (dotted line). The three ranges of ordinary turbulence are: EPR - Energy pumping range, Inertial range (IR) and Diffusion range (DR). The $q$ range relevant to the MSD in the case of ordinary turbulence, at very long times, is EPR. In that region the velocity field correlations for ordinary turbulence and for SWN turbulence are the same up to pre factors.

It seems, at first sight, that the velocity correlations of SWN turbulence have nothing to do with the velocity correlations of ordinary turbulence, in which we are interested in this article, because of the obvious difference in the properties of the driving noise as depicted in fig. (1). It turns out, however, that although the SWN model has its own physical realizations, it is also relevant to the understanding of the small $q$ correlations in ordinary turbulence. The reason is that deep in the energy pumping range, where $D(q)$ of ordinary turbulence is a constant, the driving noise does not look different from spatial white noise. (Clearly, the correlations of high momentum modes ($q > q_0$) of the velocity fields in ordinary turbulence are very different from the corresponding correlations of SWN systems. The point however is how those correlations feed into the equation describing the low momentum ($q < q_0$) correlations. It turns out that the effective low momentum theories, obtained by taking into account the different high momentum behavior of ordinary turbulence and SWN turbulence, differ only in their effective low momentum viscosities. High momentum behavior when fed into the equation for low momentum velocity correlations, only renormalizes the viscosity. The reader interested in the technical points raised here will find the discussion of a similar problem in ref. [15], where high momentum modes are separated from the low momentum modes and their effect on the low momentum modes is seen just to renormalize some constants.) Our conclusion is that the correlations of the low momentum Fourier components of the velocity field in ordinary turbulence behave in the same way as the corresponding correlations of a liquid which is driven by SWN and are thus given by equation (6), which we are going to use as the low momentum velocity field correlations for ordinary turbulence.

Since this does not complicate matters, we generalize equation (6) to

$$\Phi(q,t) = Aq^{-\Gamma} \exp[-Bqt^{1/z}] \qquad (7)$$

and the three dimensional problem to a general dimension, $d$.

(For further discussion of the stretched exponential form of the scaling function see also [16-17]. That specific form will not be essential, however, to our following considerations). It will be further assumed that the distribution governing the velocity field is Gaussian. This assumption seems adequate in two extreme cases, the case where the fluid is driven by SWN, which is the case we are interested in and the case where the fluid is driven by noise which is extremely infrared in space [14].
 (For ordinary turbulence the use of a Gaussian distribution is questionable [18, 19] but even in that case the Gaussian distribution even with $\delta$ function correlations in time, has been used to model turbulent flow [20], because of its relative simplicity and the fact that the corresponding passive scalar problem still remains rich and interesting [21, 22].)

The position vector of a particle carried by the liquid obeys the differential equation

$$\dot{\mathbf{R}} = \mathbf{V}(\mathbf{R},t). \qquad (8)$$

The probability density of a given trajectory, $\mathbf{R}(t)$ for times between $0$ and $T$ starting at the origin, is given by [21-23]

$$P\{\mathbf{R}(t)\} = < \prod_{t'=0}^{T} \delta[\dot{\mathbf{R}}(t') - \mathbf{V}(\mathbf{R}(t'),t')] >, \qquad (9)$$

where the average is taken over the noise distribution.

Introducing a three dimensional vector field $\mathbf{k}(t)$ and the standard representation of the $\delta$ function, $\delta(x) = \frac{1}{2\pi} \int_{-\infty}^{\infty} e^{ikx} dk$ and since $\mathbf{R}(t)$ is prescribed and does not depend on the specific realization of the noise, the probability density above can be written as

$$P\{\mathbf{R}(t)\} \propto \int D\mathbf{k}(t) \exp[i\int_0^T dt \mathbf{k}(t) \cdot \dot{\mathbf{R}}(t)] \left\langle \exp[-i\int_0^T dt \mathbf{k}(t) \cdot \mathbf{V}(\mathbf{R}(t),t)] \right\rangle, \qquad (10)$$

where $D\mathbf{k}(t)$ denotes functional integration over the vector function $\mathbf{k}(t)$, defined on the interval $0 \le t \le T$. The average on the right hand side of equation (10) can be performed directly to yield

$$\left\langle \exp[-i\int_0^T dt \mathbf{k}(t) \cdot \mathbf{V}(\mathbf{R}(t),t)] \right\rangle =$$
$$= \exp[-\frac{1}{2}\int_0^T\int_0^T dt_1 dt_2 \left\langle V_i(\mathbf{R}(t_1),t_1) V_j(\mathbf{R}(t_2),t_2) \right\rangle k_i(t_1) k_j(t_2)] \quad (11)$$

where the indices $i$ and $j$ denote Cartesian components and summation on double indices is assumed. The average on the right hand side of equation (11) is with respect to the noise and is an explicit function of the time difference, $t = t_2 - t_1$ and $\Delta \mathbf{R} = \mathbf{R}(t_2) - \mathbf{R}(t_1)$. The average over the noise is a tensor that on general symmetry grounds must have the form $\Lambda_{ij} = \delta_{ij}\Psi_1(|t|,|\Delta\mathbf{R}|) + (1-\delta_{ij})\frac{\Delta R_i \Delta R_j}{\Delta \mathbf{R}^2}\Psi_2(|t|,|\Delta\mathbf{R}|)$. To simplify, we average next over trajectories. This is justified since we are only interested in scaling properties of the MSD at large times and the typical distance between particle positions scales with the time difference as the average of that distance. Only the diagonal part, $\delta_{ij}\Lambda_{ii}(t)$, survives the averaging and it may be expressed as

$$\Lambda_{ii}(t) = \frac{2}{3}\int d\mathbf{q}\,\Phi(q,t) f(q,t), \quad (12)$$

where

$$f(q,t) = \left\langle \exp i\mathbf{q} \cdot (\mathbf{R}(t_2) - \mathbf{R}(t_1)) \right\rangle. \quad (13)$$

Since, $f(q,t)$ can be expressed, at least for large $t$ as $g(q^2\Delta^2(t))$, the kernel that enters into the exponent of the right hand side of equation (11), is given by

$$\Lambda_{ii}(t) = \frac{2}{3}\int d\mathbf{q}\,\Phi(q,t) g(q^2\Delta^2(t)). \quad (14)$$

For large $t$, $\Delta^2(t) \propto t^{2\lambda}$. Therefore, the integral over $q$ has a cut-off either at $q \sim t^{-1/z}$, when $1/z \geq \lambda$ or at $q \sim t^{-\lambda}$, when $\lambda \geq 1/z$. For large $t$, $\Lambda_{ii}(t) \propto t^{-\alpha}$ with

$$\alpha = [(d-\Gamma)/z]\vartheta[1/z - \lambda] + [(d-\Gamma)\lambda]\vartheta[\lambda - 1/z], \quad (15)$$

where $\vartheta$ is the step function and where it is assumed that $d > \Gamma$. The $\mathbf{k}$ functional integration can be performed leading to a probability distribution of a trajectory, $\mathbf{R}(t)$,

$$P\{\mathbf{R}(t)\} \propto \exp[-\frac{1}{2}\int d\omega \frac{\omega^2}{\phi(\omega)} \mathbf{R}(\omega) \cdot \mathbf{R}(-\omega)], \quad (16)$$

where $\phi(\omega)$ is the Fourier transform of $\Lambda_{ii}(t)$ and $\mathbf{R}(\omega)$ is the Fourier transform of $\mathbf{R}(t)$. The function $\phi(\omega)$ scales for small $\omega$ as

$$\phi(\omega) \propto \omega^{-(1-\alpha)\vartheta(1-\alpha)}. \quad (17)$$

Now, from equations (16)-(17) the average $\langle \mathbf{R}(\omega) \cdot \mathbf{R}(-\omega) \rangle$ can be easily obtained and consequently its Fourier transform, $\langle \mathbf{R}^2(t) \rangle$, which is taken to scale as $t^{2\lambda}$. This results in an equation relating $\lambda$ to $\alpha$

$$2\lambda = 1 + (1-\alpha)\vartheta(1-\alpha). \quad (18)$$

Some straight forward algebra yields $\lambda = 1/2$ in the following two cases (a) if $\Gamma \leq d - 2$ regardless of $z$ and (b) if $\Gamma \geq d - 2$ and $z \leq d - \Gamma$. Super-diffusion is found for $\Gamma \geq d - 2$ and $z \geq d - \Gamma$, where

$$\lambda = \begin{cases} 1 - (d-\Gamma)/2z & \text{for } z \leq (d+2-\Gamma)/2 \\ 2/(d+2-\Gamma) & \text{for } z \geq (d+2-\Gamma)/2 \end{cases}. \quad (19)$$

According to the above in our specific case, $d = 3, \Gamma = z = 5/3$, the swept particle super-diffuses with $\lambda = 3/5$.

An alternative numerical derivation of the above result is presented in the following, which also explicitly demonstrates that the long time behaviour of $\Delta^2$ is totally determined by the small "momentum" properties of the velocity field. The Lagrangian velocity correlations of the particle are given by $\langle \dot{\mathbf{R}}(t_1) \cdot \dot{\mathbf{R}}(t_2) \rangle = \langle \mathbf{V}(\mathbf{R}(t_1),t_1) \cdot \mathbf{V}(\mathbf{R}(t_2),t_2) \rangle$, which may look exactly like the averages discussed so far (right hand side of equation (11)) but are rather different. Previously, $\mathbf{R}(t)$ was just a **given** trajectory, for which we wanted to obtain the probability density in trajectory space. As such it did not depend at all on the velocity field. Here, $\mathbf{R}(t)$ is a solution of the evolution equation (8), giving implicitly the position of the particle as function of its initial condition and functional of the velocity field. The required Lagrangian velocity correlations involve correlations of the form

$\langle \mathbf{V}(\mathbf{q},t_1) \cdot \mathbf{V}(\mathbf{p},t_2) \exp i[\mathbf{q} \cdot \mathbf{R}(t_1) + \mathbf{p} \cdot \mathbf{R}(t_2)] \rangle$. For the infinite system the average can be decoupled exactly into a product of two averages

$$\langle \mathbf{V}(\mathbf{q},t_1) \cdot \mathbf{V}(\mathbf{p},t_2) \rangle \langle \exp i[\mathbf{q} \cdot (\mathbf{R}(t_1) - \mathbf{R}(t_2))] \rangle = 2\Phi(q,t) f(q,t) \delta(\mathbf{q}+\mathbf{p}) \quad (20)$$

which leads to the Lagrangian velocity correlations, given by $3\Lambda_{ii}(t_2 - t_1)$. A detailed discussion of the above decoupling, that was used in the past in the study of a number of problems [24-28], is presented in ref. [29] for a finite system of volume $\Omega$, where it is shown that the difference between the average of the product and product of averages is order $\Omega^{-1}$.

Equation (20) results in an integro-differential equation for the MSD,

$$\frac{d^2}{dt^2} \Delta^2(t) = 32\pi \int dq \, q^2 g(q^2 \Delta^2(t)) \Phi(q,t). \quad (21)$$

The above equation is useful, because the initial conditions, $\Delta^2(0) = \Delta_t^2(0) = 0$, are known. In order to bring out the fact that the asymptotic behaviour of the MSD depends only on the small $q$ correlations, we use as a synthetic correlation function that interpolates between the spatial white noise behaviour at small $q$ and the Kolmogorov behaviour at larger $q$,

$$\Phi(q,t) = A\{\exp(-\gamma(q/q_0)^2)(q/q_0)^{-5/3} \exp(-(q/q_0)(t/\hat{T})^{3/5}) \\ + [1 - \exp(-\gamma(q/q_0)^2)](q/q_0)^{-11/3} \exp(-(q/q_0)(t/\hat{T})^{3/2})\}. \quad (22)$$

As we saw before, the specific form for the function $g$ is not very important. The important point is the cut-off of the $q$ integration, which is introduced by it. For the numerical procedure, we need, however, a specific form for $g$ and we take it to be a Gaussian, $\exp(-q^2 \Delta^2 / 6)$.

The time scale $\hat{T}$ in the correlation function is not an independent parameter but defined by

$$\hat{T} = \frac{1}{q_0} \left( \frac{\gamma^{2/3}}{2\pi A q_0^3 (3\gamma+1)\Gamma(2/3)} \right)^{1/2}. \quad (23)$$

The dimensionless parameter $\gamma$ is introduced in order to control the size of the $q$ region, where one of the terms is more dominant. Increasing $\gamma$ makes the intermediate region reach smaller $q$'s. (As for the special form of the right hand side of the equation (23), it is enough to notice that it has the correct units.) Equation (21) is solved numerically, with the correlations defined by equations (22) and (23). Figure (2) gives the long time behavior of the MSD as obtained from that numerical solution with $\gamma = 0.0125$. The MSD is given as a function of dimensionless "time", $t/\hat{T}$. The MSD is found to be super-diffusive with $\Delta^2 \propto t^{1.22}$, in agreement with our analytical result. The behaviour of the MSD is only slightly super-diffusive and difficult to distinguish from ordinary diffusion over relatively short time scales. The smaller the energy pumping region (which is determined by $q_0/\gamma$) longer times are needed to observe super-diffusion. For the same reasons, it might not be easy to observe experimentally the super-diffusive behavior in ordinary turbulence. First, any experiment is done on a finite system, say, of linear size $L$. The size of the energy pumping region must be large enough, in order to see something. Namely, in our notation we need, $q_0 L/\gamma >> 1$. Then we have to consider the observation time. In our numerical solution we have used a natural time scale, $\hat{T}$ and saw that in order to observe super-diffusion, we had to observe the MSD for times $T >> \hat{T}$. Therefore, any experimental set up will have to take that into account. On the other hand we cannot observe the swept particle, for times in which it will reach the boundary of the system. This means that observation times, $T$ should be small enough to ensure $L/(\Delta^2(T))^{1/2} >> 1$. It has also to be taken into account, that before observation of the particle starts, we need to wait long enough until the system reaches a state, which is close enough to steady state. Essentially, it seems that although special consideration should be give to the above requirements, they can be met by having a large enough system and an energy pumping range, which does not depend on the size of the system. We expect that careful experiments designed for that purpose, will eventually lead to observation of super-diffusion in ordinary turbulence.

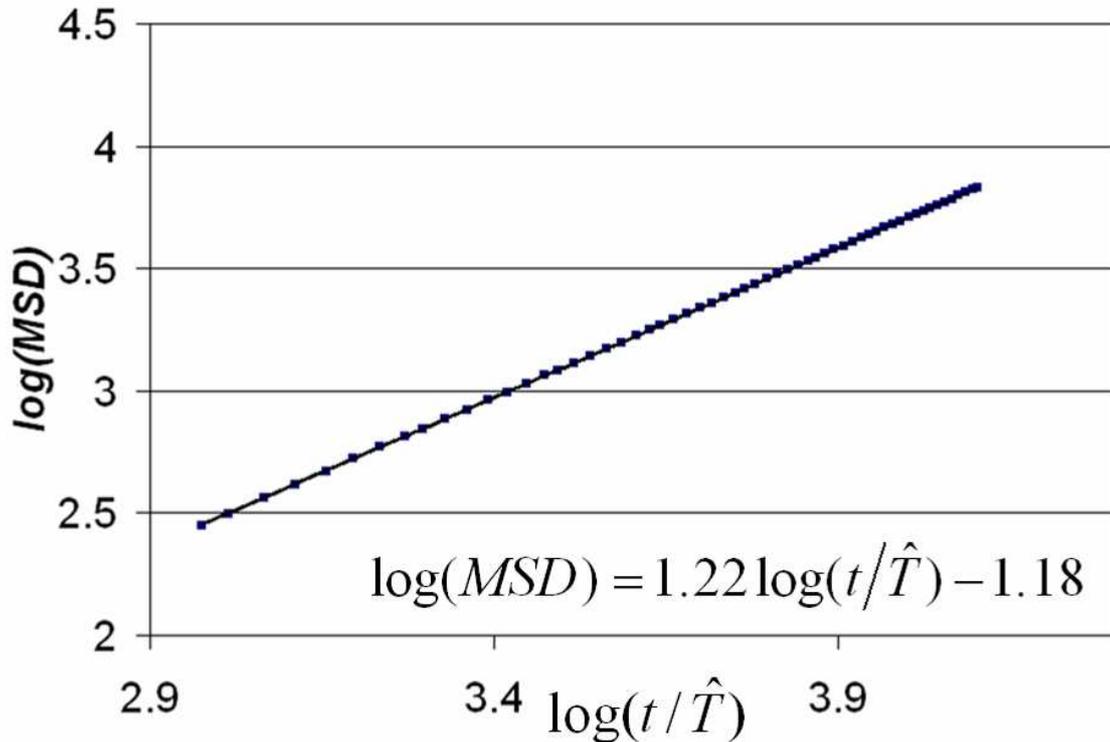

Figure 2. Long time behavior of the MSD for $\gamma = 0.0125$. The curve of $\log(MSD)$ that is obtained from Eq. (21), marked by the squares, agrees perfectly with a linear fit with slope 1.22, demonstrating super-diffusion.